# A Multisensory Approach to Probing Scattering Media

Muralidhar Madabhushi Balaji[1,2*], Danyal Ahsanullah[1] and Prasanna Rangarajan[1†]

[1]Electrical and Computer Engineering, Lyle School of Engineering, Southern Methodist University, Dallas, Texas – 75205

[2]Wyant College of Optical Sciences, University of Arizona, Tucson, AZ 85721, USA.

*mmadabhushibalaji@smu.edu

†prangara@smu.edu

## Abstract

Non-invasive detection of objects embedded inside an optically scattering medium is essential for numerous applications in engineering and sciences. However, in most applications light at visible or near-infrared wavebands is scattered by the medium resulting in the obscuration of the embedded objects. Existing methods to overcome scattering generally rely on point-by-point scanning strategies, which limit spatial sampling density. In this work, we address the sampling limitations by drawing inspiration from multisensory integration mechanisms observed in nature, wherein distinct sensing modalities work together to enhance the perception of the surroundings. Our multisensory approach leverages the unique advantages of coherent light by combining the sensitivity of an interferometric LiDAR with a wide field neuromorphic sensor to probe objects inside a densely scattering medium. The neuromorphic camera provides wide field spatial cues of the embedded object, by monitoring the fluctuations in the speckle patterns produced by tuning the laser frequency. These spatial cues are used to guide a point-scanning FMCW LiDAR to retrieve high-resolution images. Using this approach, we imaged objects embedded within an $8\ cm$ thick ($> 100$ transport mean free paths), tissue-like scattering medium with a $10\times$ improvement in sampling density compared to traditional uniform sampling.

## Introduction

In many imaging and sensing applications, the scattering of light imposes fundamental limitations on probing objects with visible and near-infrared (NIR) light. For instance, in detecting tumors within tissue[1,2], identifying diseased regions inside fruits[2-4] or navigating in foggy conditions, scattering obscures the desired information by disrupting the one-to-one correspondence between the source and detector. Consequently, functional or structural variations introduced by objects embedded within the medium cannot be accurately located, nor can their spatial and spectral features be reliably identified. Typically, this challenge is addressed by using complementary techniques that are unaffected by scattering, such as Magnetic Resonance Imaging (MRI)[5], Ultrasound[6], or Radar[7]. However, these methods are often tailored to specific applications and come with limitations that hinder their broader adoption. For instance, MRI devices are costly, bulky, and unsuitable for patients with metallic or electronic implants, while Radar or ultrasound may lack the endogenous contrast needed for applications like food inspection. Thus, there is a need to develop a strictly optical approach to detect objects embedded deep inside scattering media.

Existing optical methods address this challenge by modeling light transport as a diffusion process. This is because, over distances exceeding ten times the transport mean free path ($\ell^*$), light transport through a scattering medium begins to resemble heat diffusion. The transport mean free path, $\ell^*$ is the average length over which the direction of incident light becomes randomized due to scattering. In this diffusive regime, the embedded objects are effectively obscured by the scattering medium. Methods like Diffuse Optical Spectroscopy (DOS)[8-10] leverage diffusion-based forward models and iterative optimization algorithms to extract the optical properties i.e., scattering or absorption coefficients or dynamics of the scatterers at each voxel within the medium.

The inverse problem is inherently ill-posed and underdetermined, as the number of unknowns far exceeds the number of measurements[8]. To address this limitation and enhance reconstruction fidelity, DOS methods acquire a spatially dense set of measurements. In addition, to prevent crosstalk between measurements, DOS techniques rely on point-scanning acquisition scheme, requiring scans across numerous source-detector locations[10]. This approach significantly increases the complexity of the instrumentation and lengthens the acquisition time. Overcoming these constraints remains a key focus of ongoing research. Emerging solutions aim to increase sampling density by employing dense arrays of sources and detectors, using sophisticated opto-electronic interfaces[10-12]. These systems require complex multiplexing and synchronization schemes for data acquisition.

Rather than scaling hardware complexity, in this work, we address the challenge of limited sampling resources by drawing inspiration from the multi-sensory integration scheme seen in animals. Multi-sensory integration refers to the interdependent operation of distinct sensing schemes to enhance the perception of the surrounding environment[13]. For instance, the ability of a pedestrian to identify a hazy object on a foggy day can be significantly improved by a subsequent auditory cue, such as the sound of a horn or engine, assisting in their safe navigation[14]. Sea turtles rely on magnetic sensors for long distance navigation in an ocean, and subsequent chemical or visual cues to locate their nest on the beach[15]. Honeybees rely on visual information to approach flowers and subsequently utilize information from olfactory sensors to recognize the flower containing honey[16]. Analogously, in the multisensory integration scheme presented here, we use spatial cues obtained from a wide field system to direct a point scanning system to probe a scattering medium. This enables the detection of inhomogeneities over a wide field of regard while simultaneously identifying their spatial characteristics.

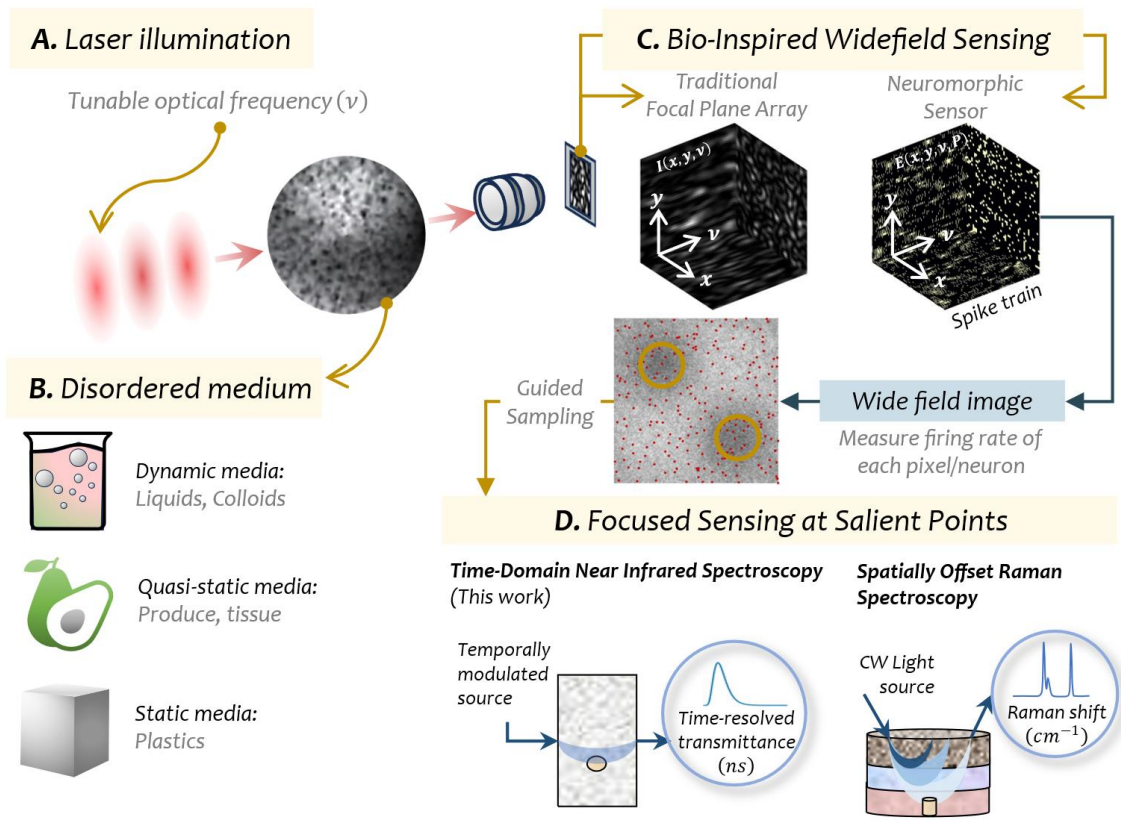

Figure 1: Proposed multisensory integration concept. The tunable laser A) illuminates a disordered medium B) to produce a temporally varying speckle pattern. The bio-inspired neuromorphic camera monitors the temporal variations in the speckle to assemble a wide-field image. C). The spatial cues obtained from the wide-field image are used to guide a point scanning approach (here, an FMCW LiDAR to recover high-resolution imagery) D).

In the proposed multi-sensory integration approach, we combine information from two distinct coherent sensing techniques that exploit the statistics of speckle patterns observed by tuning the frequency of the illumination laser. For wide field detection, we measure the rate of evolution of speckle patterns at the distal end of the medium using a highly sensitive neuromorphic camera that exclusively responds to intensity changes that can span $> 120 \ dB$ dynamic range. This approach leverages the fact that the rate of evolution of speckle with optical frequency is closely related to the local transport properties of the medium[17-23]. The neuromorphic camera converts the spatio-temporal variations in the incident light to a sequence of asynchronous events, called spike trains[24].

The mean firing rate of these spike trains at each pixel is used to generate a low-resolution image of the embedded object. The spatial cues in the recovered image are used to guide a point-scanning Frequency Modulated Continuous Wave (FMCW) LiDAR[25] based time-resolved data acquisition system. The point scanning FMCW LiDAR system is used to measure time-resolved diffuse transmittance at the locations identified from the wide field image provided by neuromorphic camera. This information is used to resolve higher-spatial frequency features on the embedded inhomogeneity. The high-dynamic range of neuromorphic sensors combined with the heterodyne gain and shot-noise limited performance of an interferometric LiDAR apparatus enables our approach to image embedded objects through an $8 \ \text{cm} \ (> 100\ell^*)$ thick static scattering medium.

## Results

*Wide field detection of embedded inhomogeneities using a Neuromorphic camera*

The presence of scattering or absorption inhomogeneities within a medium causes measurable changes in the plenoptic properties of scattered light, such as its intensity or time-of-flight distribution. By capturing these variations across a wide field of view, one can reconstruct an image of the embedded inhomogeneities. Traditionally, this is achieved by measuring the time-of-flight distribution using pulsed light sources and time-resolved detectors, such as SPAD arrays. However, these detectors often suffer from low spatial resolution and require specialized electronics to handle large data volumes. To overcome these limitations, our group recently introduced a speckle differencing approach that leverages the statistical properties of speckle patterns to image embedded inhomogeneities[23]. Speckle patterns are stochastic intensity patterns with a granular structure that are observed when coherent light illuminates a scattering medium[26]. They result from the interference of light waves with randomized phase delays caused by multiple scattering events within the medium.

In the speckle differencing approach, we leverage the fact that the cross correlation of the complex-valued speckle fields observed at optical frequencies $\bar{\nu} \ Hz$ and $\bar{\nu} + \Delta\nu \ Hz$ is proportional to the frequency response of the medium (at $\Delta\nu \ Hz$)[17-23]. Since the frequency and temporal responses (or time-of-flight distribution) are a Fourier transform pair, spatial variations in the temporal response can be inferred from speckle patterns generated under tunable laser illumination. Specifically, localized changes in the reduced scattering coefficient $\Delta\mu'_s$ and absorption coefficient, $\Delta\mu_a$ of the inhomogeneities cause variations in the temporal response, which are encoded in the speckle patterns. We further identified that the temporal differences between speckle patterns, acquired under tunable laser illumination, directly encode information about these variations. This insight enables the use of neuromorphic sensors that record only differential changes, significantly reducing data throughput.

Building on this insight, we developed a neuromorphic sensing-based approach to detect inhomogeneities embedded within a scattering medium. Neuromorphic sensors act as fast change detectors, asynchronously generating "spikes" or "events" in response to time-varying stimuli incident on their pixels[24]. Each pixel, independently generates a "spike" whenever the natural log intensity $(\Delta \log I)$ changes by a value that exceeds a preset threshold $(\Theta_{on}\ or\ \Theta_{off})$, as shown in the simulations shown in Figure 2. The number of events generated over a fixed duration correlates with the rate of intensity fluctuations. Consequently, the total number of events serves as a proxy for the spectral correlation width and, by extension, the scattering properties of the medium. This capability enables neuromorphic sensors to operate effectively under rapidly changing conditions while significantly reducing data volume. Furthermore, their logarithmic response and exclusive sensitivity to intensity fluctuations allow them to operate in scenes with a dynamic range exceeding $120\ dB$.

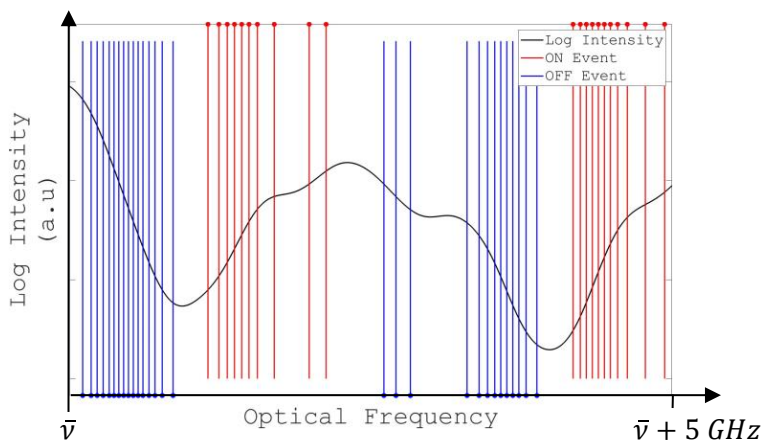

A. *Simulated spectral speckle for a scattering medium with $L/\ell^* = 12$*

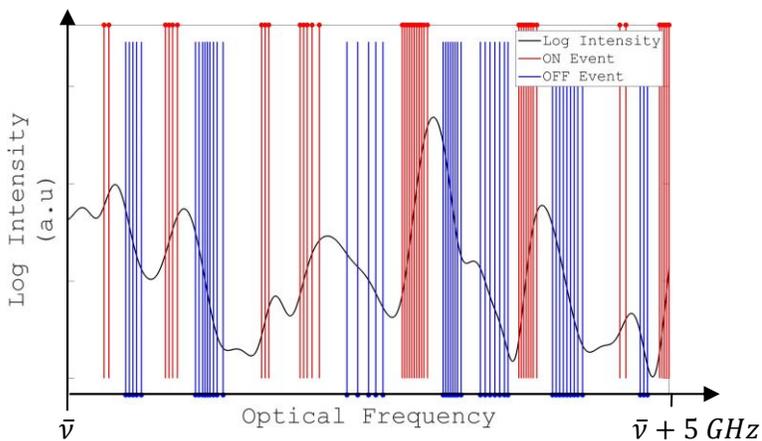

B. *Simulated spectral speckle for a scattering medium with $L/\ell^* = 100$*

Figure 2: Simulated speckle patterns and the corresponding response of a neuromorphic pixel for A) $12\ell^*$ thick medium and B) $100\ell^*$ thick medium. The rate of speckle fluctuation varies with the scattering thickness. Simulated using the approach described in Ref. 23

These unique advantages have driven their adoption across diverse imaging applications where high dynamic range and sensitivity are critical[24,27-29].

In our approach, rapid tuning of the laser emission frequency induces speckle fluctuations. The evolution rate of these patterns is governed by the local time-of-flight distribution, which is perturbed by embedded inhomogeneities, leading to detectable changes in speckle dynamics. We adopt this approach to obtain wide field images of inhomogeneities embedded within a scattering medium as shown in Figure 3A. The computational procedure described in the Methods is utilized to tune the emission frequency at a rate of $2\ GHz/s$ for $400\ GHz$. A laser beam at $780\ nm$ (Methods) was expanded to a diameter of $150\ mm$ to illuminate the scattering medium. A Neuromorphic sensor (Prophesee EVK4) monitors the resulting time-varying speckle pattern at the distal end of the scattering medium of $60l^*$ thickness (Section S1).

The inhomogeneities were introduced into the scattering medium by inserting matte paper cutouts of different shapes. The positive and negative events triggered by the neuromorphic camera during the sweep were accumulated and subsequently denoised using a $3 \times 3$ median filter to generate the wide-field images of inhomogeneities (see Figure 3C). A comparison of the neuromorphic imaging approach with a traditional imaging approach is additionally provided in Section S2. The sensitivity of the neuromorphic approach to detecting changes is determined by the threshold parameters $(\Theta_{on}\ and\ \Theta_{off})$ and the sweep rate $(\alpha)$ used to generate the events. These parameters were optimized empirically as described in Section S3. The resulting wide-field images provide spatial cues to guide a point-scanning, time-resolved data acquisition system described in the next section.

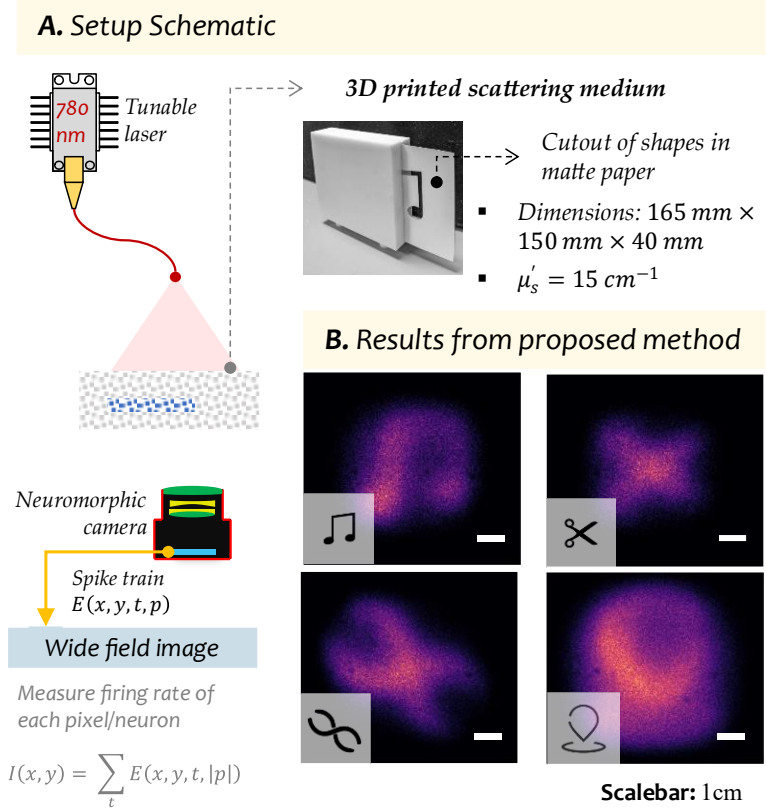

Figure 3: A) Schematic of the setup used to experimentally retrieve spatially resolved images of inhomogeneities embedded inside a scattering medium of scattering thickness of $60l^*$. B) Obtained results for different inhomogeneities.

*Acquiring the time-of-flight distribution of the medium using FMCW LiDAR*

The time-of-flight distribution of light is also commonly referred to as the temporal point spread function (TPSF) of the medium. The TPSF of a scattering medium is akin to the impulse response of an LTI system that fully characterizes the behavior of the system. Therefore, TPSF measurements contain valuable information that can be used, for instance, to characterize the transport properties of the medium[30] discern the scattering contributions from absorption inhomogeneities[31,32,33] or even improve the spatial resolution of images of embedded objects[34]. In this section, we demonstrate the use of FMCW LiDAR by characterizing the transport properties of media with distinct scattering properties. Subsequently, we showcase the integration of FMCW LiDAR with the neuromorphic camera system to identify the spatial features of the embedded object.

The FMCW LiDAR technique was originally developed for 3D ranging applications, where it is used to measure the round-trip time-of-flight of light from objects in the scene. These techniques allow the detection of ultrafast temporal changes in the light distribution, occurring at GHz – THz bandwidths, by downmixing them into kHz to MHz range[35]. Moreover, the heterodyne gain in an interferometric apparatus can be used to perform shot-noise limited data acquisition. Both these factors allow the FMCW methods to measure ultrafast temporal variations with high sensitivity without needing specialized equipment such as high bandwidth, single photon sensitive detectors. These advantages prompted the development of several FMCW LiDAR techniques for the purpose of recovering time-resolved information from a scattering medium[36-42].

The FMCW LiDAR uses a Mach Zehnder interferometric arrangement shown in Figure 4A. The light in the reference and probe beams are sourced from a narrow linewidth tunable laser. Typically, the angular frequency varies linearly at a rate $\alpha = \Delta v(T^{-1})$, where $\Delta v\ Hz$ is the total sweep range and $T\ secs$ is the total sweep duration. The probe beam is used to illuminate the scattering medium. The scattering medium causes the light to scatter and emerge at different time delays relative to the reference beam. Each time delayed wavefront, delayed by $\tau_i$, interferes with the reference beam to produce a distinct beat signal, oscillating at a frequency of $(\alpha \tau_i)\ Hz$ (Section S4) at the detector[25]. The power spectrum of the measured beat note provides an estimate of the time-resolved transmittance or reflectance of the medium.

The temporal resolution of the FMCW LiDAR system is inversely proportional to the total linear sweep range $\Delta v\ Hz$ (Section S4). However, non-linearities in the optical frequency sweep introduce phase fluctuations that deteriorate the temporal resolution. Therefore, methods like active sweep compensation of large bandwidth tunable lasers[43], stitching the frequency sweeps from multiple lasers[44], or post processing and resampling procedures[45] are employed to enhance the temporal resolution of these systems. We adopted an inverse filtering-based frequency sweep linearization technique here to linearize the frequency sweep of the tunable laser (Methods).

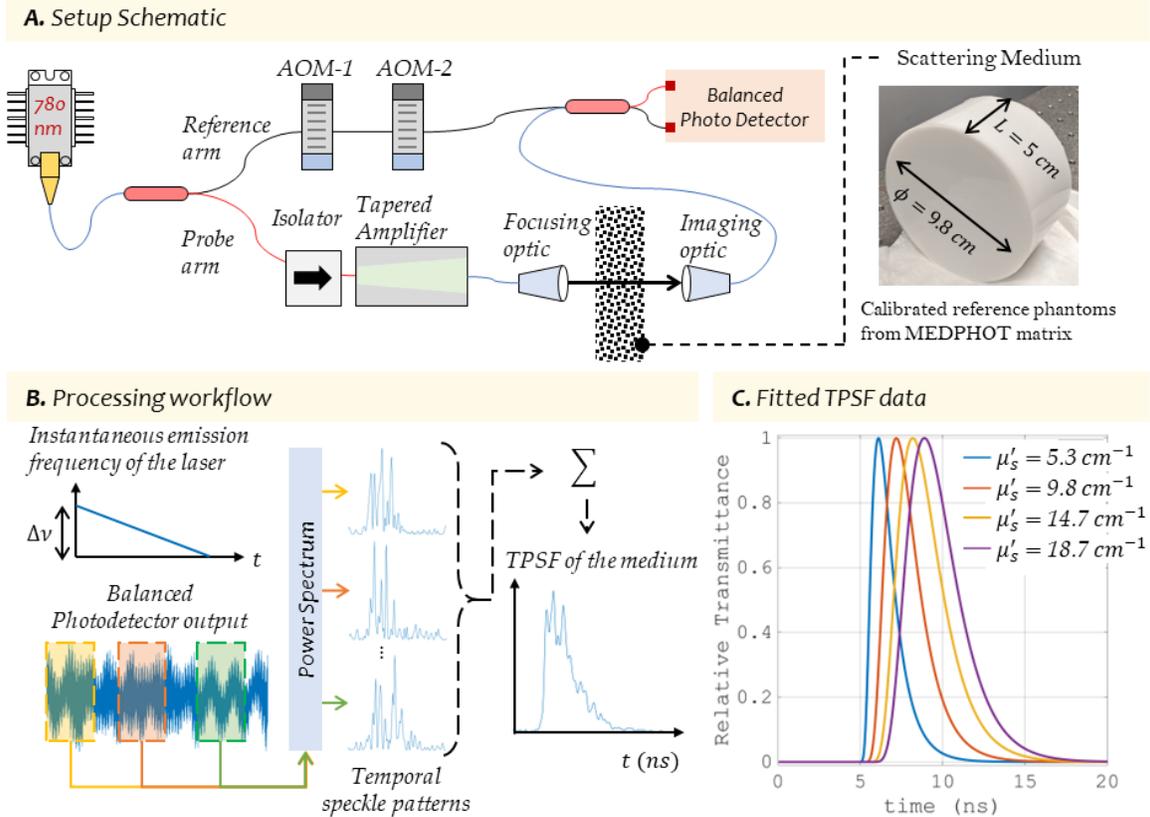

Figure 4: A) Schematic of the setup used to acquire the TPSF of a scattering medium using FMCW LiDAR B) Processing workflow and C) Fitted TPSF's for media of different scattering properties.

The measured time-resolved transmittance is additionally influenced by speckle arising from the interference of randomly phased wavefronts. Although the statistics of time-resolved speckle contain information that can be used to extract gas concentrations[39]. or blood flow[42], they pose a hindrance when estimating the transport properties. Typically, the microscopic motion of scatterers produces distinct temporal speckle patterns, which are then averaged to reduce speckle noise[39,41,42]. Due to the absence of microscopic motion in the static scattering samples considered here, we adopted a strategy inspired by Bartlett's or Welch's method of averaging periodograms[46]. In this approach, we subdivide a single interferogram into multiple segments equal length, but randomized time offsets. Each interferogram is additionally multiplied by a window function. The resulting independent speckle realizations in the power spectrum averaged to minimize the effect of speckle noise.

We used the experimental setup shown in Figure 4A to estimate the transport properties of scattering phantoms chosen from a MEDPHOT matrix[47] (see Methods). We acquired 121 measurements for each phantom by scanning the phantom in a square grid with $0.5\ mm$ inter sample spacing. These sample points are situated close to the central region of the phantom. Figure 4B illustrates the sub-division process described earlier, wherein the experimentally acquired interferogram is sub-divided into multiple windows of fixed length. The window size was chosen to realize a temporal resolution of $45\ ps$. The power spectrum of each window is computed to produce a time-resolved speckle pattern, as shown in Figure 4B. Each time-resolved speckle pattern is then fit to an analytic expression of the TPSF using a non-linear least squares routine (Section S5). The fitted TPSF's corresponding to different scattering coefficients are shown in Figure 4C. It can be noticed that, as the scattering increases, the peak of the TPSF shifts to the right, indicating a delay in the time of arrival, which is consistent with theory[30]. Additional statistical details of the measurements are provided in Section – S6 of the supplementary document. In the next section, we describe how this system is integrated with a neuromorphic sensing approach to enable imaging through a $100\ell^*$ thick scattering medium.

*Multisensory integration to image through $100\ell^*$*

The proposed multisensory approach begins with probing the scattering medium using a neuromorphic sensing-based technique to obtain a wide-field image of inhomogeneity. This image is then used to spatially guide the point-scanning FMCW LiDAR for acquiring time-resolved transmittance measurements from the scattering medium. This integration of information from distinct sensing modalities mirrors the multi-sensory integration paradigm observed in biological systems. The schematic workflow of our approach is shown in Figure 5B. To demonstrate this method, we use an $8\ cm$ thick static scattering medium of $\mu_s' = 15\ cm^{-1}$ and $\mu_a = 0.01\ cm^{-1}$, resembling those of breast tissue. The incident light travels through a length that corresponds to $120\ l^*$ through the medium, far exceeding the cases considered in Figure 3. Both sensing modalities operate in a transillumination geometry. Within this medium, we embedded two scattering inhomogeneities made from paraffin wax (see Section S1), as shown in Figure 5A.

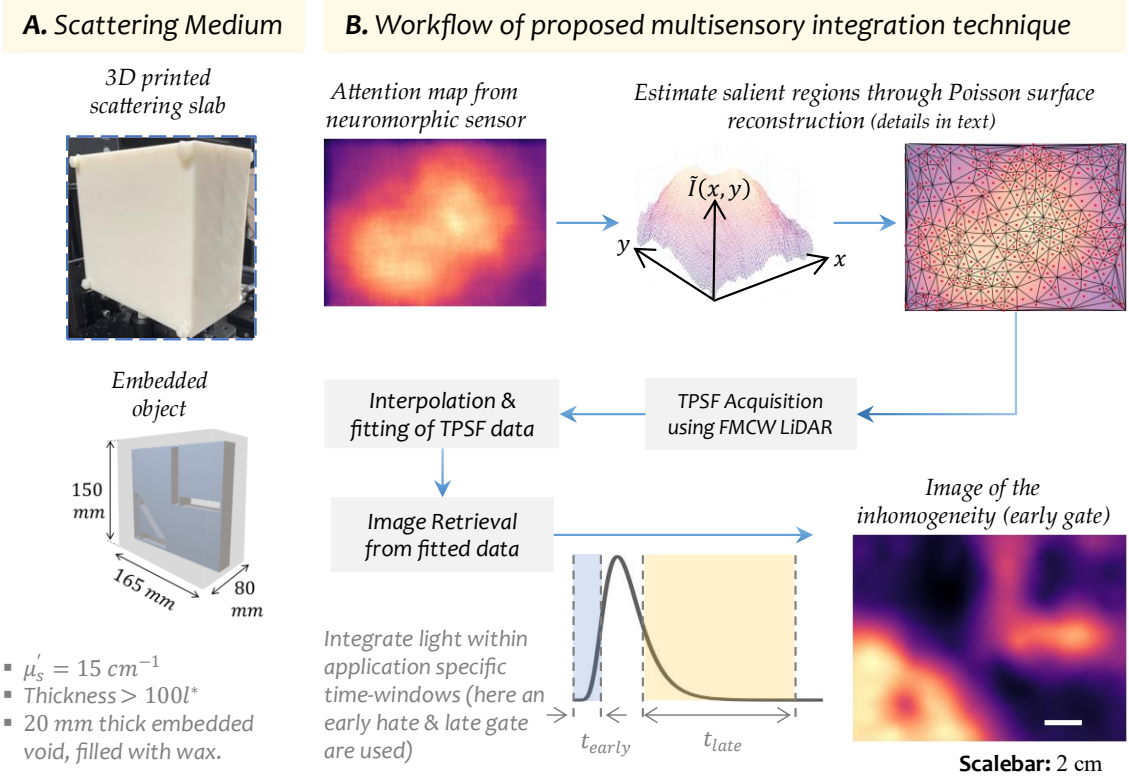

Figure 5: Proof of principle demonstration of the proposed multisensorial integration approach. A) 8 cm thick static Scattering medium used in the experiments. B) Workflow of the approach to recover images of embedded inhomogeneities. C) Comparison with images obtained using a conventional camera under LED illumination in a transillumination geometry (Section S3).

Using the neuromorphic sensor approach described earlier, we obtained a wide-field image of the embedded inhomogeneities. The emission frequency of the laser was tuned by varying the injection current with a triangular waveform. Although this approach detects the presence of embedded objects, the resulting image lacks sufficient spatial detail (see Figure 5B). This is attributed to the use of area illumination in the neuromorphic sensing method, which introduces crosstalk between adjacent spatial pixels in the wide field image[48]. Nonetheless, the wide-field image provides valuable cues by highlighting regions with high information content. To extract these cues, we use a Poisson surface reconstruction (PSR) procedure[49] to identify the salient sampling points where time-resolved transmittance measurements can be acquired.

The PSR framework reconstructs the surface of the object by exploiting the integral relationship between the point cloud data and the scalar indicator function $\chi$ that defines the 3-dimensional object. The scalar indicator function $\chi$ typically takes the value of 1 for all the points inside the surface and a value of 0 for points outside. Thus, the gradient of the indicator function, $\nabla\chi$ represents the surface normal at each point on the point cloud and has a non-zero magnitude only near the surface of the object. Furthermore, the measured point cloud, $\vec{P}$ represents the gradient of the indicator function, $\nabla\chi$. Thus, the Poisson Surface Reconstruction framework solves an optimization problem to find the indicator function $\chi$ whose gradient $\nabla\chi$ closely approximates the point cloud measurements $\vec{P}$.

This procedure generates a refined mesh that best fits the point cloud surface, with the density of triangles in the mesh determined by the local curvature of the reconstructed surface. Consequently, the density of triangles indicates regions exhibiting significant variations transport properties. We identify the vertices or incenters of these triangles and acquire time-resolved photon scattering function (TPSF) measurements at these locations using the FMCW LiDAR. In our experiments, the scattering medium is scanned to acquire TPSF at the identified locations on the surface. These measurements were interpolated and fit to an analytic model to recover a dense spatio-temporal stack of TPSF measurements (Section S7). The images the inhomogeneous regions can be assembled the estimated spatio-temporal data stack, using various techniques, such as displaying the $\mu_s'$ and $\mu_a$ values obtained through fitting[50,51], calculating temporal moments[52], or by temporal integration of data over specific time windows[31-33,53]. Here, we employ a temporal integration approach, to recover the images of the inhomogeneities. This approach is used in transillumination geometry to discriminate between scattering and absorption inhomogeneities[31-33]. Similar procedures have also been used in reflection geometry to obtain depth-selective images of inhomogeneous regions[53]. Here, we perform a time integration of the fitted spatio-temporal stack over an early time window to retrieve image of scattering inhomogeneity.

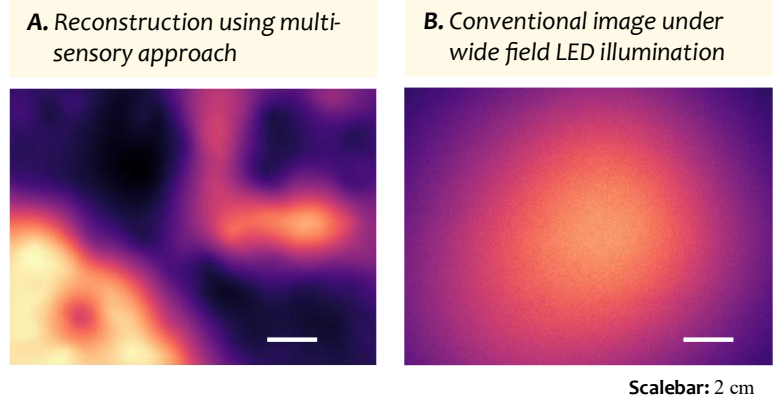

Figure 6: Comparison of the reconstructions obtained using A) proposed approach with B) images obtained under conventional wide field LED illumination.

**Scalebar:** 2 cm

To compare the improvement in the ability to reconstruct the embedded inhomogeneity, we acquired an incoherent image of light distribution seen at the distal end of the scattering medium. For this, an LED operating at a center wavelength of 850 nm was homogenized and expanded to illuminate an 80 mm diameter on the scattering medium. The intensity distribution in a transillumination geometry was captured using a camera (additional details in Section S2). The results shown in Figure 6, demonstrate the advantage of using our approach to recover high frequency spatial features. In addition, having access to time-resolved data allows us the ability to discern scattering inhomogeneities from absorptive inhomogeneities. We show this capability in the experiments outlined in Section S8, where the time-resolved data integrated over early and late time windows reveal the scattering and absorption inhomogeneities, respectively.

It is worth noting that the PSR procedure employed here acts as an importance sampling mechanism by prioritizing regions of the point cloud that exhibit strong surface gradients. This effectively concentrates sampling on areas with significant geometric features, improving the likelihood of retrieving meaningful time-resolved data. However, alternative importance sampling strategies could also be adopted depending on the specific imaging goals. For instance, entropy-based methods could be used to target areas with higher informational uncertainty, or clustering algorithms like DBSCAN or k-means could segment the wide-field image into regions of interest based on spatial features. Each of these alternatives aims to reduce the total number of sampling points required by focusing resources on the most informative regions of the scene, similar to how PSR leverages surface cues from the point cloud.

# Discussion

We presented a bio-inspired multi-sensory integration framework designed to optimize the allocation of sampling resources for imaging through densely scattering media. By combining the complementary advantages of interferometric detection and neuromorphic sensing, we demonstrate the ability to probe media with scattering thicknesses $> 100\ell^*$. The proposed multisensory approach enables the recovery of fine spatial details of embedded objects without the need for dense source-detector arrays and complex hardware synchronization. In the results shown in Figure 5, this method achieved a minimum inter-sample spacing of ~1.5 mm while requiring only 437 measurements in total. In comparison, a conventional uniform sampling grid at the same resolution would require ~6400 measurements, reflecting an order of magnitude reduction in total measurements.

This resource-efficient strategy is particularly advantageous in scenarios where acquisition speed is critical, such as functional brain imaging[9], where rapid localization of neuronal activity is essential. Moreover, adopting such multisensory framework to existing diffuse imaging modalities like Spatially Offset Raman Spectroscopy (SORS)[54] offers a compelling direction for future research. Techniques like SORS often lack precise spatial information about the location of subsurface inhomogeneities. Integrating spatial guidance from a complementary sensing modality, as demonstrated in our framework, could significantly improve targeting efficiency, and reduce unnecessary sampling.

Although we chose a static scattering medium here, the methods described here can also be extended probe quasi-static or dynamic scattering media (see Section S9). For instance, a faster frequency tuning mechanism such as drive current modulation can be employed to acquire time-resolved measurements from dynamic media. Moreover, existing neuromorphic cameras also possess the ability to detect high-speed changes making them suited for probing dynamic media. These cameras, with their—high-speed *(≈ 1μs resolution)*, dynamic range *(>120 dB)*, thresholding mechanism, pixel density *(≈ 1 megapixels),* and low data throughput—have the potential to unlock novel capabilities in applications that rely on speckle dynamics[55-57]. Alternative sensing modalities like asynchronous SPAD arrays that rely on a similar strategy can also prove useful for such applications requiring high speed of operation and low data throughput. The results shown here also demonstrate that the 1-bit spike trains encode sufficient information to detect embedded objects. Therefore, these sensors can potentially be used to circumvent the high data transfer rate challenges often seen in imaging through dynamically scattering media.

However, it must be noted that, despite the microsecond timestamp resolution, the detection speed of neuromorphic sensors is constrained by the latency of readout circuitry. Each event generated within the pixel-array is processed on a "first-come-first-served" basis through an arbitration mechanism[27]. If multiple pixels trigger events simultaneously, the read-out circuitry can become overwhelmed, leading to increased latency and reduced acquisition speed. Thus, the tradeoff between the spatial density of the events and the detection speed requires careful consideration in applications like DCS. Additionally, the sensitivity and event generation rate of a neuromorphic pixel depend on various threshold settings (bias parameters). In Section S6, we describe a procedure outlined to optimize these settings, similar to methods proposed in the literature. However, more systematic approaches for determining these parameters are currently under investigation by several research groups.

While our approach of using accumulated spatial events to generate a wide-field image from spike trains was effective, there is potential for further refinement. Advanced techniques like spiking neural networks[58] could leverage the full spatio-temporal statistics of spike trains, potentially enhancing sensitivity and detection speed. These networks are designed to process temporal data more efficiently, mimicking the way biological neurons communicate. Our approach—relying exclusively on endogenous contrast mechanisms at non-ionizing wavebands—is particularly suited for applications such as food inspection, where the use of exogenous contrast agents or ionizing radiation presents challenges.

## Methods

### Experimental setup

The FMCW LiDAR measurements were acquired using the setup schematic shown in Figure 4A. The output from the tunable DBR laser (Photodigm PH780DBR040BF-ISO, linewidth < 1 MHz) was split into a reference arm and a probe arm using a polarization-maintaining fiber optic waveguide splitter. The light in the reference arm is passed through a cascade of acousto-optic modulators (AOMs) that sequentially upshifts (Brimrose TEF-110-10-60-780-2FP-PM-HP+) and downshifts (Brimrose TEF-110-10-60-780-2FP-PM-HP-) the optical frequency to produce a net frequency shift of 110 kHz. In the probe arm, the optical power was amplified using a tapered amplifier (Toptica BoosTA Pro, 780 nm) to generate 180 mW of optical power. The amplified beam was focused on to a 0.5 mm diameter spot on the scattering medium. The scattered light at the distal end of the medium was imaged on to one of the two input ports of a fiber optic coupler (KS Photonics, CTDC-P-078-1-FA). The fiber coupler was used to interfere the scattered light with the light in the reference arm. The resulting interferogram was recorded on an adjustable gain balanced photoreceiver (Newport 2107).

### Scattering media

The 3D-printed phantoms utilized in our experiments were fabricated using a mixture of white pigment (RS-F2-CRWH-01) and Color Base resin (RS-F2-GPCB-01), both commercially available from Formlabs. We adhered to the standard recipe and mixing protocol prescribed by Formlabs to synthesize the white resin and print the targets. The scattering phantoms from the MEDPHOT matrix were procured from BioPixs (Trace code: BioPixS0030). The transport properties of these phantoms were characterized by the vendor through time-domain reflectance spectroscopy (provided in Section S6).

### Realizing linear sweeps through thermal tuning

Our approach requires a narrow linewidth laser source whose tunable frequency range exceeds $500\ GHz$. For this purpose, we employ a Distributed Bragg Reflector (DBR) semiconductor laser (Photodigm PH780DBR040BF-ISO). The emission frequency of the laser we used could be tuned over $\approx 1\ THz$ through temperature modulation. As the temperature of the laser changes, the refractive index within the cavity varies, altering the lasing mode and consequently the emission frequency. However, thermal modulation results in a slow and non-linear optical frequency sweep. To address this, we have developed an inverse filtering-based method that improves the linearity and speed of these sweeps. This technique uses the impulse response of the laser to determine the optimal modulation waveform, enabling rapid and linear optical frequency sweeps.

In our work, the temperature of the laser is controlled using a Thermo-Electric Cooler (TEC) integrated within the laser. The TEC consists of an array of semiconducting PN junctions sandwiched between two conducting surfaces. An electric current is established through the PN junctions to transfer the heat between the two conducting surfaces, through a process known as the Peltier-Seebeck effect[59]. The heat transfer in the lasing medium occurs through multiple intermediary layers, making the structure and thermal conductivities of these layers critical for determining the thermal tuning characteristics of the laser. Analytically, the thermal impulse response of the DBR laser can be modeled as a summation of exponential functions, with time constants dependent on the thermal responsivities of the laser's components[60,61].

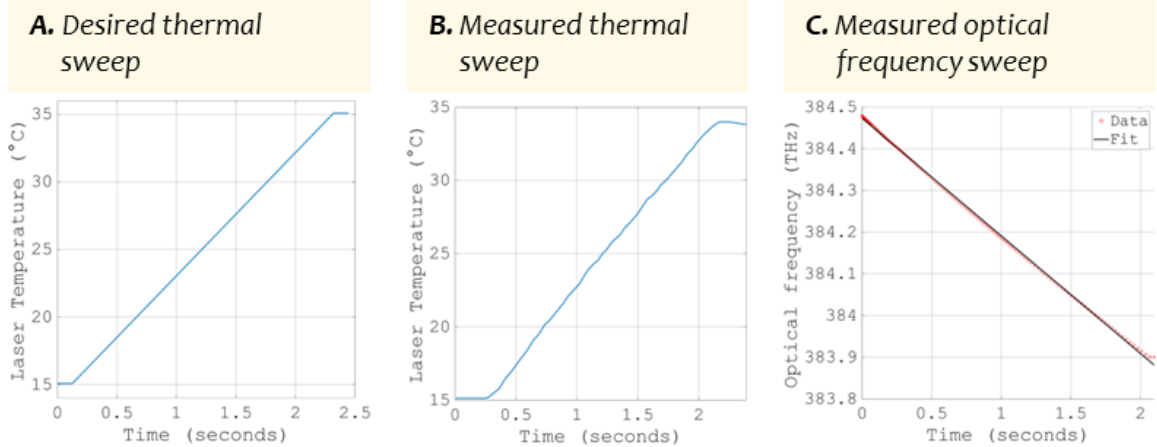

Figure 7: A) The desired thermal sweep B) Measured thermal sweep using the package thermistor and C) The instantaneous emission frequency of the laser corresponding to the temperature sweep.

The inverse-filtering procedure we employ relies on accurately estimating the time constants of the impulse response of the laser. To achieve this the step response of the laser, i.e., the temperature response in the DBR laser, to a step change in the input TEC current is measured. A 0.5A step input is applied using a four-quadrant current source, and the resulting thermal transient is measured using a thermistor integrated within the laser. The measured step response is then fitted to an analytic expression of the thermal step response of a DBR laser[60,61], represented as weighted sum of exponential functions,

$$T_{step}(t) = \sum_{i=1}^{N} a_i \exp\left(-\frac{t}{\tau_i}\right) \qquad (1)$$

For our laser, we empirically chose $N = 2$, to fit the measured step response. The fitting procedure yielded the time constants of $\tau_1 = 10\ seconds$ and $\tau_2 = 700\ seconds$. These values were used to computationally generate the impulse response of the laser. The impulse response, and the desired temperature sweep were input to the *filter()* function in MATLAB to obtain the desired modulation waveform. Additionally, the modulation waveforms generated by the inverse filtering procedure were constrained to the safe operating limits of the TEC, restricting the maximum achievable sweep rate. Through this approach, the highest achievable thermal sweep rate was determined to be $10\ °C/second$.

The computational inverse filtering approach is validated by applying the current modulation waveform to the TEC using a four-quadrant current source. The resulting temperature profile of the laser was measured by recording thermistor readings with a Vescent SLICE-QTC temperature controller. The desired temperature profile (Figure 7A) and the measured laser temperature profile (Figure 7B) show close agreement. Additionally, the measured waveform, $9.5\ °C/second$, obtained by fitting the data to a first-degree polynomial closely matches the desired thermal slope of 10 °$C/second$. The residual error is attributed to transient, second-order effects in the laser and TEC behavior that are not modeled here. The instantaneous wavelength of the laser, corresponding to the optimized modulation waveform, was also measured using a wavemeter (*High finesse, WS-8*), and is shown in shown in Figure 7C. The wavelength sweep was fitted to a first-degree polynomial, estimating a sweep rate of $280\ GHz/s$, and a sweep excursion of $\approx 560\ GHz$. Additional details of the validation of the approach using the FMCW LiDAR system are provided in Section S10.

*Poisson surface reconstruction*

An open-source library, known as Open3D is used in this work to identify the spatial locations in the wide field image with high information content. In this procedure, the normalized wide-field image is viewed as a point cloud, with the 2D pixel locations acting as the $(x, y)$ coordinates and the normalized intensity at the pixel as the $z$ coordinate. A 3D surface that best fits the point cloud data is discovered by Open3D using a Poisson Surface Reconstruction framework. This procedure finds a smooth indicator function $\chi$ from which the iso-surface is identified as the points where $\chi = 0.5$, i.e., mid-way between the interior and the exterior of the object. A Delaunay triangulation algorithm is then used to extract a surface mesh of triangular elements that are formed by a 3D triangulation of sample points on the smooth surface. The sample points are refined, and more points are added to the mesh iteratively to satisfy the size and shape criterion of the triangular elements. The procedure outlined above, implemented in Open3D is used to extract a triangular surface mesh, comprising vertices, edges and faces that define the 3D surface. The above procedure ensures that the regions exhibiting large surface height variations are represented using more triangles or vertices. Thus, the vertex density in the surface mesh is an indicator of the spatial gradient distribution, and thereby a proxy to the information content in the wide field image. Therefore, the subsequent FMCW LiDAR data is acquired at these vertex locations. The time resolved FMCW LiDAR measurements were acquired at these locations by scanning the scattering sample.

## Acknowledgments

**Funding:** This work was supported in part by DARPA under agreement number HR00112290052. The U.S. Government is authorized to reproduce and distribute reprints for Governmental purposes notwithstanding any copyright notation thereon. Any opinions, findings and conclusions or recommendations expressed in this material are those of the authors and do not necessarily reflect those of the sponsor. M.M.B. was supported by the Dean's Dissertation Fellowship from the Moody School of Graduate and Advanced Studies, Southern Methodist University, and gratefully acknowledges this support.

**Author contributions:** Conceptualization: P.R. and M.M.B. Methodology: P.R., M.M.B. and D.A. Investigation: M.M.B Visualization: P.R. and M.M.B. Supervision: P.R. All authors contributed to editing the manuscript. We used ChatGPT 4.0 to enhance the clarity and coherence of the manuscript. We reviewed and revised the content as necessary and take full responsibility for the publication's content.

**Competing interests:** All authors declare that they have no competing interests.

# Supplementary Material for
## A Multisensory Approach to Probing Scattering Media

Muralidhar Madabhushi Balaji *et al.*

*mmadabhushibalaji@smu.edu, prangara@smu.edu

## S1. Optical properties of the scattering media

The optical properties of the scattering media used in this work are detailed below. The 3D printed phantoms used in our experiments were produced by mixing white pigment (RS-F2-CRWH-01) with Color Base resin (RS-F2-GPCB-01), both commercially available from Formlabs. We followed the standard recipe and mixing protocol provided by Formlabs to synthesize the white resin and print the targets. The refractive index and absorption coefficient of Formlabs clear resins were reported to be $\mu_a \approx 0.1\ cm^{-1}$ and $n \approx 1.46$ for $750\ nm - 800\ nm$ waveband in Ref.1.

To estimate the reduced scattering coefficient ($\mu_s'$) of our phantoms, we use the Oblique Incidence Reflectometry (OIR)[2]. In OIR, the scattering medium is illuminated with a pencil thin beam, incident at an oblique angle, $\theta_{inc}$. From the definition of a transport mean free path ($l^*$), the incident light propagates a distance of $\approx 1l^*$ along the refracted path before its direction is completely randomized. Consequently, the randomized or diffuse light emerging from the medium is spatially shifted relative to the incident beam, as shown in Figure S1. The spatial shift ($\Delta x$), incidence angle ($\theta_{inc}$), and refractive index of the medium ($n$) can be used to determine the transport mean free path of the medium, using the expression,

$$\mu_s' \approx \frac{\sin(\theta_{inc})}{n\Delta x} \quad (2)$$

We 3d printed a homogeneous block of dimensions $110\ mm \times 110\ mm \times 50\ mm$ to measure the $\mu_s'$ of the resin. A laser beam is focused onto the scattering medium to illuminate it at a $60°$ angle. Due to the significantly lower intensity of the diffuse light, we acquired two images: one with a low exposure setting to measure the location of the incident beam, and another with a high exposure setting to measure the location of the backscattered beam. The pixel shift along the central row was measured from these two images. The median pixel shift was then multiplied by the magnification of the imager to estimate the shift ($\Delta x$) in physical units. From this, the $\mu_{s,resin}'$ is estimated to be $\approx 15.6\ cm^{-1}$.

For the wax samples used as inhomogeneities, the refractive index and absorption coefficient were reported to be $\mu_a \approx 0.04\ cm^{-1}$ and $n \approx 1.46$ in the $750\ nm - 800\ nm$ waveband[3]. We used the same procedure described above to estimate the reduced scattering coefficient. A laser beam was focused onto a wax block at an angle of 60°. The median pixel shift was multiplied by the magnification of the imager to estimate the physical shift ($\Delta x$) in physical units. From this, the $\mu'_{s,wax}$ is estimated to be $0.1\ cm^{-1}$.

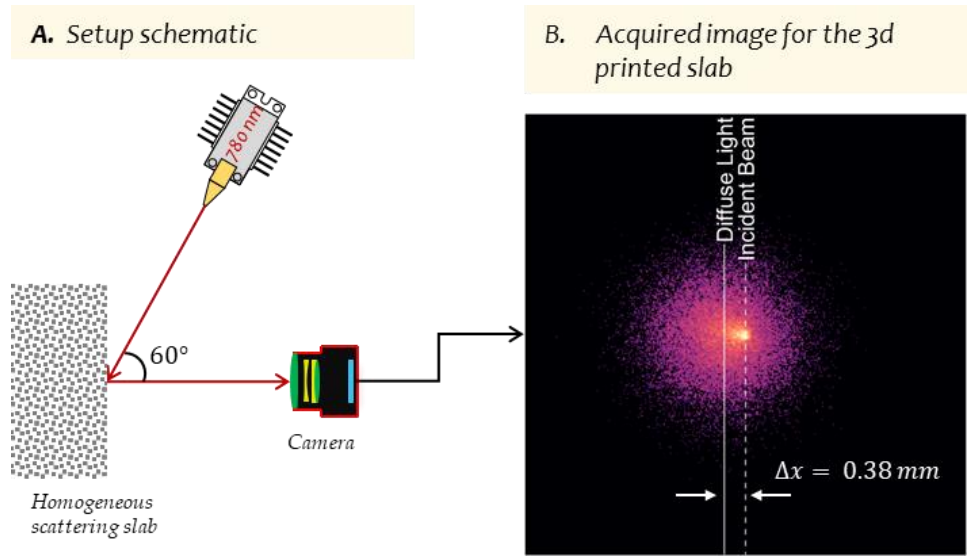

Figure S1: Oblique incidence reflectometry to characterize the scattering properties of the medium. A) Setup schematic and B) the image of the incident and diffuse light distribution acquired by the camera

## S2. Neuromorphic imaging approach vs Traditional imaging

In this section, we provide a comparative assessment of the images obtained using the neuromorphic approach and conventional imaging. The experimental details of the neuromorphic sensing approach are provided in the main manuscript. A similar setup was used to acquire intensity images under LED illumination. The LED emission (Smart Vision Lights, ODSXP30-850) at 850 nm was homogenized and expanded to illuminate an 80 mm diameter on the scattering medium. The intensity distribution in a transillumination geometry was relayed to a camera (IDS UI-3000SE) using an imaging optic (SIGMA 35mm F1.4 DG HSM).

The intensity images of the objects embedded within the scattering medium (the same as those used in the main manuscript) were acquired using the above setup. The field of view of the intensity camera was approximately matched to that of the neuromorphic camera. It is evident from the results shown in Figure S2 that the neuromorphic approach offers a significant enhancement in the detectability of the embedded objects compared to the traditional imaging approach.

## S3. Impact of bias parameters on the output of neuromorphic sensor

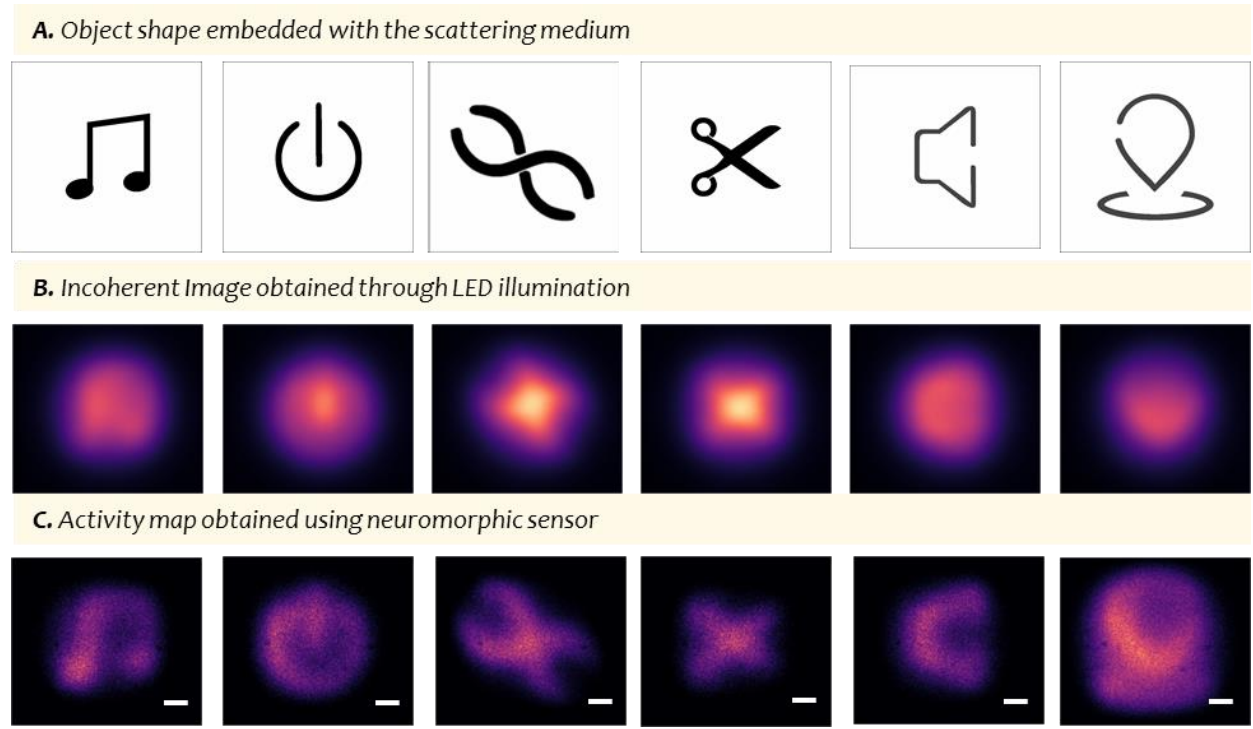

Figure S2: A) Embedded object B) Incoherent intensity images acquired under a traditional imaging setup C) Images assembles using the neuromorphic camera based setup.

The sensitivity of the neuromorphic approach in detecting an inhomogeneity is determined by the threshold parameters $(\Theta_{on}\ and\ \Theta_{off})$ and the sweep rate $(\alpha)$ used to generate the events. By adjusting these parameters, the spiking behavior on the event sensor can be optimized to enhance the contrast in the activity map. The selection of ON and OFF thresholds $(\Theta_{on}\ and\ \Theta_{off})$ for event generation impacts the noise levels and temporal contrast sensitivity of an event sensor. A lower threshold increases the sensitivity of the event sensor to temporal contrast variations but introduces more noise. Conversely, a higher threshold reduces the sensitivity of the event sensor temporal contrast variations but reduces the noise events. This tradeoff must be carefully assessed while setting the threshold values of the event sensor.

This trade-off becomes particularly relevant when considering large $\Theta_{on}$ or $\Theta_{off}$ the events are generated only at regions that exhibit large temporal variations in intensity. For such thresholds, events are generated only in regions with significant temporal intensity variations. This mechanism can be used to tailor event generation to focus exclusively on areas with large speckle variations caused by inhomogeneities, thereby improving their detectability. However, caution is required to ensure that the thresholds do not exceed the highest temporal contrast in the scene, as this would result in the event sensor producing no events.

Similarly, the sweep rate of the laser determines the rate at which the speckle fluctuates at the detector. As we have discussed in the main manuscript, the detection speed of a neuromorphic camera is constrained by the latency of readout circuitry. Thus, the number of events that can be generated without overwhelming the sensor depends on the sweep rate of the laser. These two parameters must be carefully adjusted to obtain the best performance from a neuromorphic sensor.

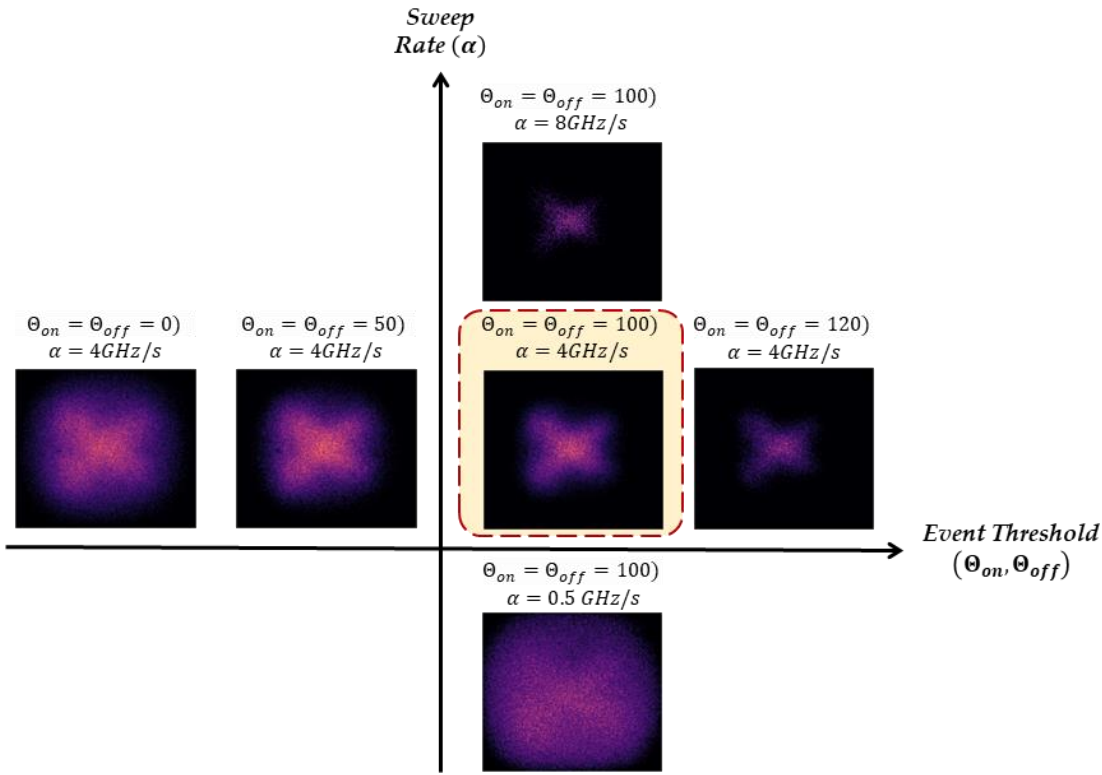

Figure S3: Impact of bias parameters and sweep rate on the accumulated events map obtained from neuromorphic sensor for an exemplar target.

In this work, we adopted an empirical procedure, whereby the threshold and sweep rate values that produce the best visual output was chosen for data acquisition. The images were recorded for a range of threshold and sweep rate values, as shown in Figure S3. The ON and OFF thresholds ($\Theta_{on} = \Theta_{off} = 100$) and a sweep rate of $4\ GHz/s$ that produces the highest contrast, while still producing signal events are used in the experiments shown the manuscript. Although a rudimentary procedure shown here produced good results, more elaborate procedures that consider the impact of all threshold parameters on the generated events can be used to improve the results[4,5,6].

## S4. FMCW LiDAR model

In an FMCW LiDAR, the output from a narrow linewidth tunable laser is split into a reference beam and a probe beam, as shown in Figure S8. The time delayed copies of the probe beam that emerge after interacting with the object are collected and interfered with the reference beam at the detector. Typically, the instantaneous angular frequency of the laser is linearly modulated in an FMCW interferometer. Therefore, the instantaneous angular frequency, $\omega_R(t)$ of the laser can be expressed as,

$$\omega_R(t) = \omega_0 + 2\pi\alpha t \tag{3}$$

where, $\alpha = \Delta v(T^{-1})$ is the rate of modulation, given as the ratio of the total sweep range $\Delta v\ Hz$ to the total sweep duration $T\ seconds$. The instantaneous phase of the reference beam can then be expressed as,

$$\phi(t) = \left(\int_0^t \omega_R(t)\ dt + \phi_0\right) = (\pi\alpha t^2 + \omega_0 t + \phi_0) \tag{4}$$

where, $\phi_0$ represents the initial phase of the light in the reference arm. Using Eq.(4), the expression for the optical field in the reference beam is obtained as,

$$U_R(t) = U_0 \exp(j(\pi\alpha t^2 + \omega_0 t + \phi_0)) \tag{5}$$

where, $U_0$ is the amplitude of the reference beam which is assumed to remain constant over the sweep duration. Similarly, the time-delayed copies of the probe beam reaching the detector can be expressed as,

$$U_P(t) = \sum_{i=1}^N U_i \exp(j(\pi\alpha(t-\tau_i)^2 + \omega_0(t-\tau_i) + \phi_0)) \tag{6}$$

where, $N$ denotes the number of time-delayed copies of the incident beam reaching the detector. The spatial variations in the amplitude are ignored for brevity. The instantaneous intensity at the detector due to the interference between the reference beam $U_R(t)$ and the probe beam $U_P(t)$ is then obtained as,

$$I(t) = I_0 + \sum_{i=1}^N I_i \cos(2\pi\alpha\tau_i t + \omega_0\tau_i - \pi\alpha\tau_i^2) \tag{7}$$

From the expression for the interferogram disclosed in Eq.(7), it is evident that the interference between the reference beam and a probe beam with a time delay $\tau_i$ results in a sinusoidally oscillating intensity waveform at a frequency.

$$f_i = \alpha \tau_i \; Hz \tag{8}$$

Thus, the relative time delays of the light emerging from the medium can be estimated by computing the power spectrum of the detected beat note. To obtain the relationship between the temporal resolution of the FMCW approach and the sweep parameters, compute the derivative of Eq.(8),

$$\Delta f = \alpha \Delta \tau \tag{9}$$

Here, $\Delta \tau$ denotes the temporal resolution of the FMCW approach. The frequency resolution ($\Delta f$) can further be described using the relationship $\Delta f = f_s N^{-1}$, where $f_s$ is the DFT sampling resolution and $N$ is the window size. By substituting the expression for $\Delta f$ and the rate of modulation $\alpha = \Delta \nu (T^{-1})$, in Eq.(9), we get the expression for $\Delta \tau$ as,

$$\Delta \tau = \left(\frac{1}{\Delta \nu}\right) \times \left(\frac{f_s T}{N}\right) \tag{10}$$

However, the window size $N$, can be expressed in terms of $f_s$ and $T$ as $N = f_s T$. Thus, the expression for the temporal resolution $\Delta \tau$ becomes,

$$\Delta \tau = (\Delta \nu)^{-1} \tag{11}$$

Eq.(11) shows that the temporal resolution of an FMCW LiDAR system is inversely proportional to the total sweep range $\Delta \nu \; Hz$. Thus, tunable lasers with tuning ranges exceeding $1 \; THz$ are required for realizing FMCW LiDAR systems with $sub-picosecond$ temporal resolution.

## S5. Fitting the TPSF to estimate transport properties

This section details the key steps involved in the fitting procedure used to estimate the transport properties from the TPSF measurements.

The time-resolved intensity, $I_{slab,meas}(t)$ obtained using the FMCW LiDAR has an unknown time offset, $t_0$ arising from the path length difference between the reference and probe beams. To estimate this time offset, we replace the scattering medium with a thin acrylic slab of $3 \; mm$ thickness. The peak value of the measured TPSF is used as $t_0$ to obtain the time shifted intensity, $I_{slab,meas}(t - t_0)$. Additionally, the measured time-resolved intensity is blurred by the finite temporal resolution of the FMCW LiDAR. We account for this in the fitting procedure by convolving the current estimate obtained from the analytic TPSF $I_{slab,DE}(L, t)$ with a Gaussian kernel with a $\sigma = 22 \; ps$ to obtain $\tilde{I}_{slab,DE}(L, t)$. The transport properties ($\mu_s'$ and $\mu_a$) of the scattering medium are obtained by minimizing the error between $I_{slab,meas}(\tau)$ and the analytic TPSF, $\tilde{I}_{slab,DE}(L,t)$, disclosed in Eq. (2) by solving a constrained optimization problem,

$$\min_{(\mu_s',\mu_a)} \left\| \tilde{I}_{slab,DE}(L,t) - \alpha I_{slab,meas}(t-t_0) \right\|^2 \tag{12}$$

The parameter $\alpha$ accounts for the unknown scale factor between the predictions of the diffusion equation and the interferometrically obtained intensity. In our fitting procedure, we empirically determine the value of $\alpha$ to be $2 \times 10^{12}$. The fit parameters were randomly initialized in range [0.01,0.05] for $\mu_a$, and between 25% and 175% of the actual value for $\mu_s'$. Additionally, we only consider regions in the TPSF with intensity value larger than 5% of the maximum intensity of the TPSF are used for computing the loss function. This approach minimizes the influence of noise on the fitting procedure and aligns with the methodology adopted in (60). The measured TPSF (average of the 121 measurements) and the fitted TPSF shown in Figure S4, can be seen to be in close agreement.

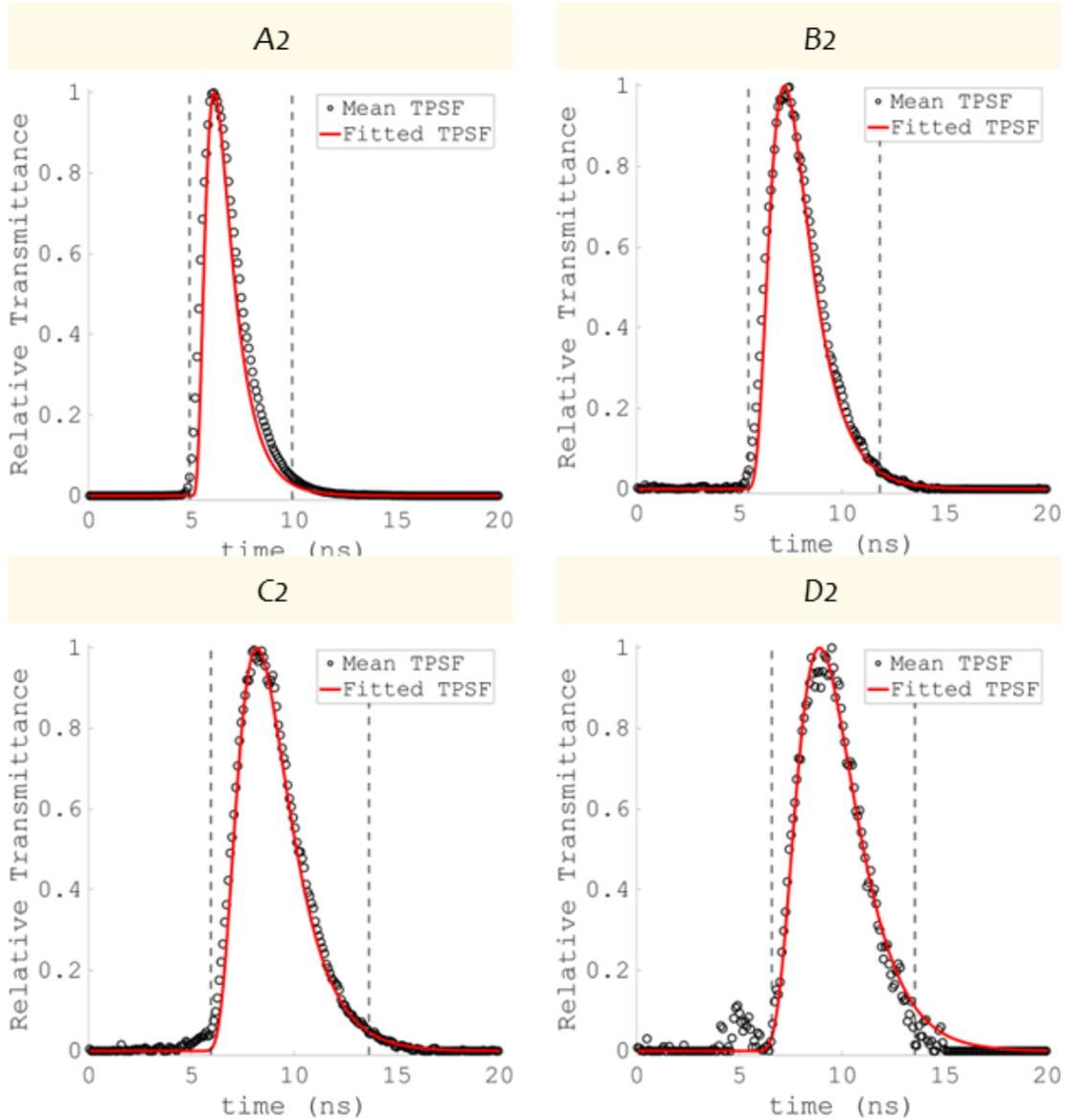

Figure S4: Measured and Fitted TPSFs acquired for the different scattering phantoms. The regions in between the vertical dashed lines are used for fitting.

## S6. Measured transport properties

We used the experimental setup shown in Figure 4A of the main manuscript to estimate the transport properties of scattering phantoms chosen from a MEDPHOT matrix. We acquired 121 measurements for each phantom by scanning the phantom in a square grid of $0.5\ mm$ inter sample spacing. These sample points are situated close to the central region of the phantom. The distinct speckle realizations produced from the sub-divided interferogram windows are averaged to obtain $I_{slab,meas}(t)$ the TPSF of the scattering medium. The TPSF is fitted to the analytic expression of the time-resolved transmittance of a slab model to estimate its optical properties. Here, we use the analytic expression of the TPSF of a scattering slab[7-9], disclosed below,

$$I_{slab,DE}(L,t) = \sum_{m=-\infty}^{m=\infty} \left\{ (L - z_{1,m}) \exp\left(-\frac{(L-z_{1,m})^2}{4Dc't}\right) - (L - z_{2,m}) \exp\left(-\frac{(L-z_{2,m})^2}{4Dc't}\right) \right\} (4\pi Dc')^{-\frac{3}{2}} t^{-\frac{5}{2}} \exp(-\mu_a c't) \times \quad (13)$$

where, $c'$ is the speed of light in the medium, $D = 1/(3\mu_s')$ is the diffusion coefficient and $L$ denotes the thickness of the medium. The distance $z_{1,m}$ and $z_{2,m}$ denote the locations of virtual sources used to solve the diffusion equation using extrapolated boundary conditions, and are defined as,

$$z_{1,m} = 2m(L + 2z_e) + z_0$$
$$z_{2,m} = 2m(L + 2z_e) - 2z_e - z_0 \quad (14)$$

where, $z_0 = 1/(\mu_s')$ is the transport mean free path of the medium and $z_e$ follows the definition disclosed in Refs 8 and 10. For the fitted TPSF's shown here, we use $m = -4\ to +4$. The fitting procedure minimizes the mean-squared error between the analytic TPSF and the measured TPSF by iteratively updating the scattering and absorption coefficients using a non-linear least squares routine (Section S10). The fitted TPSF's corresponding to variations in scattering coefficient and scattering coefficient are shown in shown in Figure 4C of the main manuscript. The predicted values and the actual values for four different scattering phantoms are provided in Table 1.

The error in estimating $\mu_s'$ is within the same range as the traditional time domain instruments (Table – 2 in Ref.11). However, the approach introduces a larger error in estimating $\mu_a$, despite predicting the correct trend. This large error in absorption is also observed in other FMCW techniques and has been attributed to the effect of windowing in Ref.12. Furthermore, traditional TD approaches overestimate absorption and underestimate scattering, the FMCW approach tends to do the reverse. Explaining the source of both these effects requires further analysis that is beyond the scope of this manuscript. The confidence interval values in Table 1, also indicate an increase in the spread of the estimated properties as the severity of scattering increases. This is potentially caused by the reduced signal-to-noise ratio in the measurements.

Table 1: Transport properties of the phantoms estimated using the proposed method.

| | | MEDPHOT Phantom code | | | |
|---|---|---|---|---|---|
| | | **A2** | **B2** | **C2** | **D2** |
| $\mu_s'$ ($cm^{-1}$) | True value | 5.3 | 9.8 | 14.7 | 18.7 |
| | Predicted value (median) | 4.9 | 11.4 | 18.8 | 25 |
| | Confidence interval $[0.25, 0.75]$ | $[4.8, 5]$ | $[11.2, 11.5]$ | $[18.5, 19.1]$ | $[24.4, 26.2]$ |
| | Relative error | $-7.5\%$ | $16.3\%$ | $27\%$ | $32\%$ |
| $\mu_a$ ($cm^{-1}$) | True value | 0.05 | 0.05 | 0.05 | 0.05 |
| | Predicted value (median) | 0.0194 | 0.023 | 0.024 | 0025 |
| | Confidence interval $[0.25, 0.75]$ | $[0.019, 0.02]$ | $[0.022, 0.024]$ | $[0.023, 0.025]$ | $[0.023, 0.027]$ |

## S7. Interpolation and Fitting

We assemble a dense three-dimensional stack of spatio-temporal data by interpolating the TPSF measurements acquired using the FMCW LiDAR. For this purpose, we use the Kernel Density Weighting (KDW) based interpolation scheme, wherein, the interpolated intensity value at the point of interest is obtained as the weighted average of the intensities at the measured sample points, $\boldsymbol{p}_i$. The interpolation scheme is an adaptation of the kernel density estimation method, which is widely used to estimate the probability density function of a random variable, as illustrated in Figure S5.

In this approach, a Gaussian kernel $K\left(\frac{\boldsymbol{p}-\boldsymbol{p}_i}{\sigma}\right)$ is defined around each measured sample point $\boldsymbol{p}_i$. The interpolated intensity value at the point of interest is obtained by summing the product of the measured intensity and the value of the Gaussian kernel $K\left(\frac{\boldsymbol{p}-\boldsymbol{p}_i}{\sigma}\right)$ evaluated at the point of interest, $\boldsymbol{p}$ i.e.,

$$I(\boldsymbol{p}, t) = \frac{1}{N} \sum_{i=1}^{N} I(\boldsymbol{p}_i, t) \ K\left(\frac{\boldsymbol{p}-\boldsymbol{p}_i}{\sigma}\right) \qquad (15)$$

The spread of the Gaussian function, denoted as $\sigma$ and also known as the bandwidth, determines the smoothness in the interpolated data. A larger bandwidth results in a smoother representation, while a smaller bandwidth produces data with well-resolved features. By utilizing information from all measured points in computing the interpolated value, the KDW interpolation scheme also helps reduce speckle noise. The reduction in speckle noise can be upper bounded by a value inversely proportional to the square root of the number of samples used in the interpolation process.

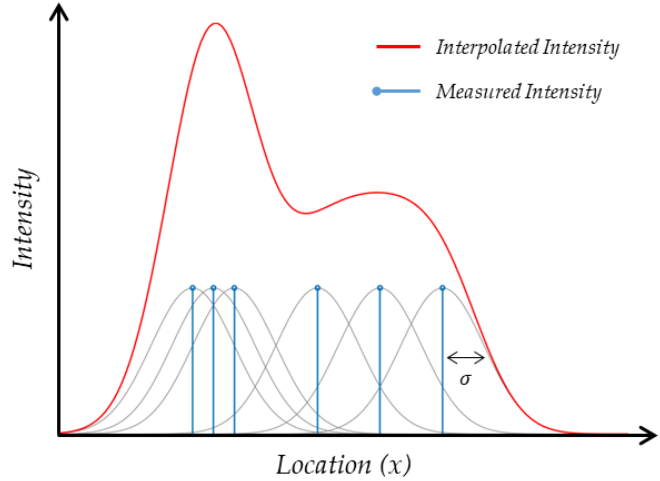

Figure S5: Illustration of Kernel Density Weighting scheme.

Despite this averaging, residual noise remains in the interpolated spatio-temporal data. Therefore, each TPSF measurement within the data stack is fitted to an analytic model loosely based on Green's function of the TPSF of a homogeneous medium, as shown below

$$I = t^{-p} \exp\left(-at - \frac{b}{t}\right) \quad (16)$$

The parameters $a$ and $b$ are related to the absorption and scattering coefficients of the medium. To obtain initial estimates for $p, a$ and $b$, a linearized version of Eq.(16) is derived by taking the logarithm on both sides. This results in a linear system of equations, where the $\log(I)$ is expressed as a linear combination of $\log(t), -t$ and $-t^{-1}$ with coefficients $p, a$ and $b$. This system of equations is solved using the *robustfit()* function in MATLAB to obtain a consensus estimate of $\hat{p}, \hat{a}$ and $\hat{b}$. These values are used as a starting guess for fitting the analytic model to each TPSF through a constrained optimization procedure implemented using *'fmincon()'* function in MATLAB. The fitting procedure aims to minimize the mean-squared error between the acquired measurements and the predicted TPSF.

The early arriving and late arriving temporal regions in the fitted data stack are identified as the time bins in the consensus TPSF (assembled from $\hat{p}, \hat{a}$ and $\hat{b}$), where the rate of change of intensity is at its maximum in the positive and negative directions, respectively. Although more sophisticated analytic models exist to represent the TPSF of light emerging from a slab-like phantom, a simpler model is adopted here for ease of implementation. A more accurate model can be used to obtain quantitative information such as the reduced scattering coefficient $\mu_s'$ and $\mu_a$ at each spatial position.

## S8. Multi-sensory integration: additional results

The workflow depicted in Figure 5 of the main manuscript was employed to recover images of scattering and absorption inhomogeneities simultaneously present in a medium. The structure of the embedded object is analogous to the Shepp-Logan phantom commonly used in CT systems, as shown in Figure S6A. In this figure, the white regions denote scattering inhomogeneities (composed of paraffin wax), and the black regions represent absorption inhomogeneities (composed of black resin). The attention map from the neuromorphic sensor identifies the sampling points for the FMCW LiDAR system, as illustrated in Figure S6B.

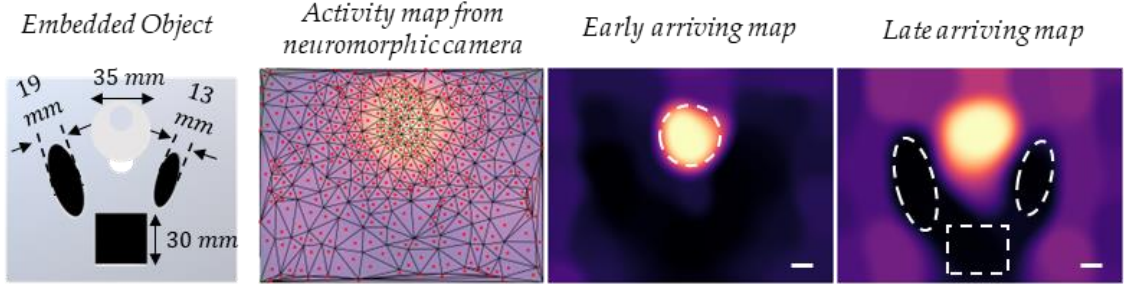

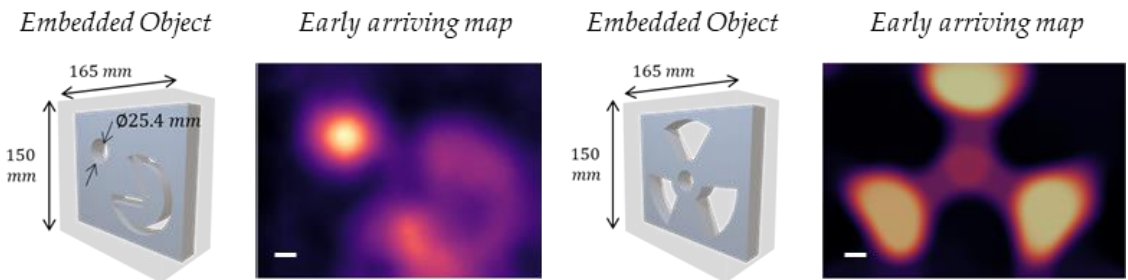

Figure S6: A Discriminating the scattering and absorptive regions using the multi-sensory approach B) Additional results using scattering inhomogeneities of different shapes. Scalebar: 1 cm

The TPSFs recorded using the FMCW LiDAR were processed using the previously outlined procedure to recover both the early arriving topographic map and the late arriving topographic map. These images were further deblurred using a total-variation-based deconvolution routine and are displayed in Figure S6A. The images demonstrate that the two temporal windows enable the discrimination of scattering and absorptive regions embedded within the medium. Additional results using scattering inhomogeneities of different shapes are provided in Figure S6.B.

**S9. Wide-field detection of inhomogeneities within realistic media**

In this section, we demonstrate the application of a wide-field neuromorphic sensing approach to detect inhomogeneous regions within realistic media such as fruit. Unlike the scattering media discussed in the main manuscript, media like fruit are quasi-static, producing time-varying speckle patterns known as "bio-speckle"[13,14]. The statistics of these bio-speckle patterns can provide valuable information about the freshness and moisture content of fruit. By analyzing the spatial variability in the temporal evolution of bio-speckle patterns, we can identify inhomogeneous regions within agricultural produce. This capability can potentially aid in detecting moldy core diseases in apples or brown heart disease in pears, facilitating the development of rapid, non-invasive food quality screening devices.

Here, we present the results of a proof-of-principle demonstration of the neuromorphic sensing approach's ability to detect inhomogeneous regions in quasi-static media. In this experiment, chayote squash serves as the quasi-static scattering medium, with the seed region acting as the inhomogeneity. As illustrated in Figure S7, light from the tunable laser is expanded to illuminate the scattering medium, and the laser is tuned using the procedure described in the main manuscript. A neuromorphic sensor detects the light distribution emerging from the medium in a trans-illumination geometry, clearly revealing the regions of inhomogeneity through the attention maps. In another demonstration to detect concealed objects rectangular regions were cutout from a black paper and positioned in between two soap bars, as shown in Figure S7B. These rectangular regions can be seen from the activity maps produced by accumulating the events generated by the neuromorphic sensor.

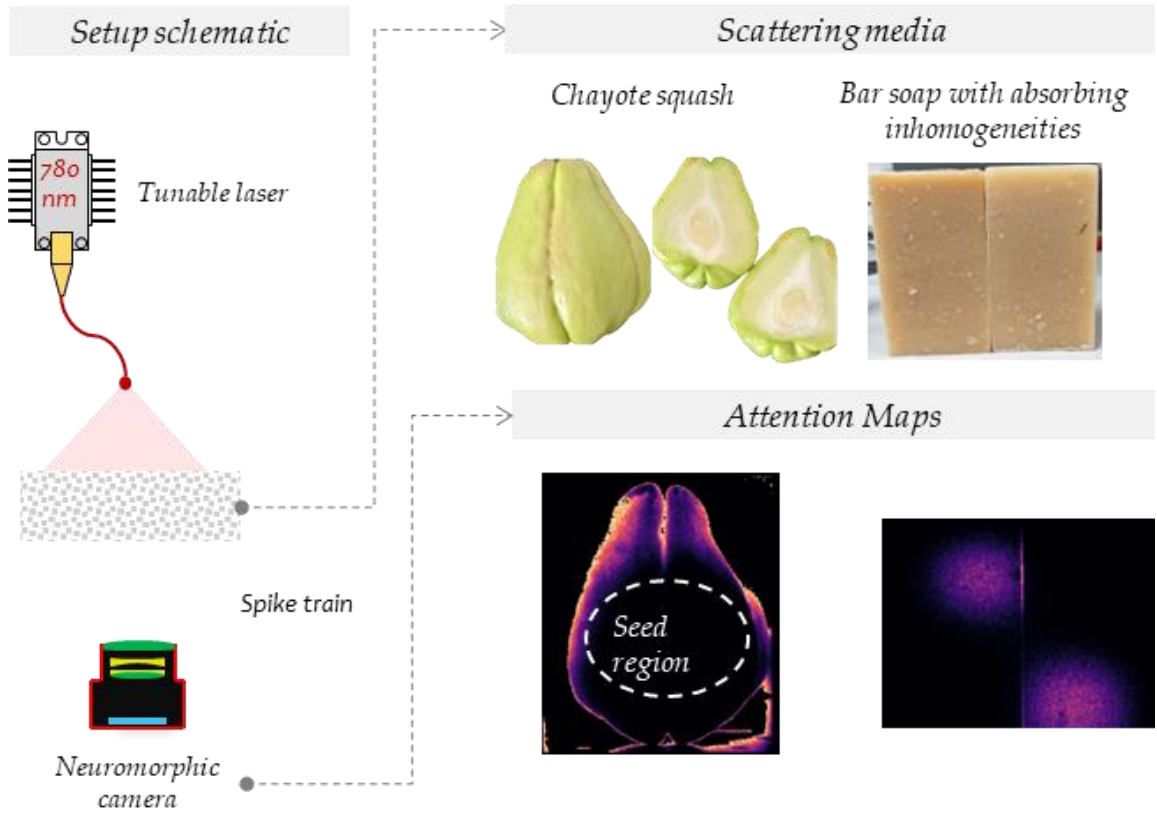

Figure S7: Detecting inhomogeneities within realistic media

## S10.　Additional details on the FMCW LiDAR experimental setup

The FMCW interferometer depicted in Figure S8 was utilized for the experiments described in the main manuscript. The laser output from the tunable laser source was directed through a series of polarization-maintaining fiber optic waveguide splitters into both a reference interferometer and the FMCW LiDAR used for probing scattering media. The reference interferometer was employed to correct for any residual nonlinearities in the optical frequency sweeps achieved through thermal modulation.

In the reference interferometer, the light in one arm passes through a series of acousto-optic modulators (AOMs) that sequentially upshift (Brimrose TEF-110-10-60-780-2FP-PM-HP+) and downshifts (Brimrose TEF-110-10-60-780-2FP-PM-HP-) the optical frequency to produce a frequency shift of 110 kHz. The optical power in the two arms is matched using a variable attenuator. A 3-dB fiber optic coupler is used to combine the light two arms and direct it towards a balanced photo receiver (Resolved Instruments, DPD80). A high-speed digitizer (GaGe) was used to digitize the output of the balanced photodetector, recording $2 \times 10^6$ samples of the interferogram at $1\ MS/s$ sampling rate. The AOM's and the fiber attenuators in the reference arm introduce a path or time delay between the two arms of the interferometer. The sweep rate ($\alpha$), determined from the wavemeter measurements, was used to calculate the time delay ($\tau$) using Eq.(8). The time delay was found to be $\approx 23\ ns$.

The residual non-linearities in the optical sweep were measured by leveraging the relationship between the instantaneous phase of the interferogram and the optical frequency sweep. Specifically, the instantaneous phase $\varphi(t)$ of the interferogram is related to the instantaneous optical frequency sweep of the laser, $\nu(t)$ and the time delay, $\tau$, through the relation $\varphi(t) = 2\pi\tau\nu(t)$ [15,16]. Thus, estimating the $\varphi(t)$ provides access to the frequency sweep of the laser. To estimate $\varphi(t)$, we demodulate the interferogram around the desired beat frequency, and apply a low pass filter the signal using 'filtfilt()' function in MATLAB. The phase of this signal is estimated using a two-quadrant arc tangent function, $atan2(...)$, followed by an unwrapping operation. The knowledge of $\tau$ was used to convert the measured phase into optical frequency, $v_n(t)$. Here, $v_n(t)$ denotes the non-linearities in the optical frequency sweep.

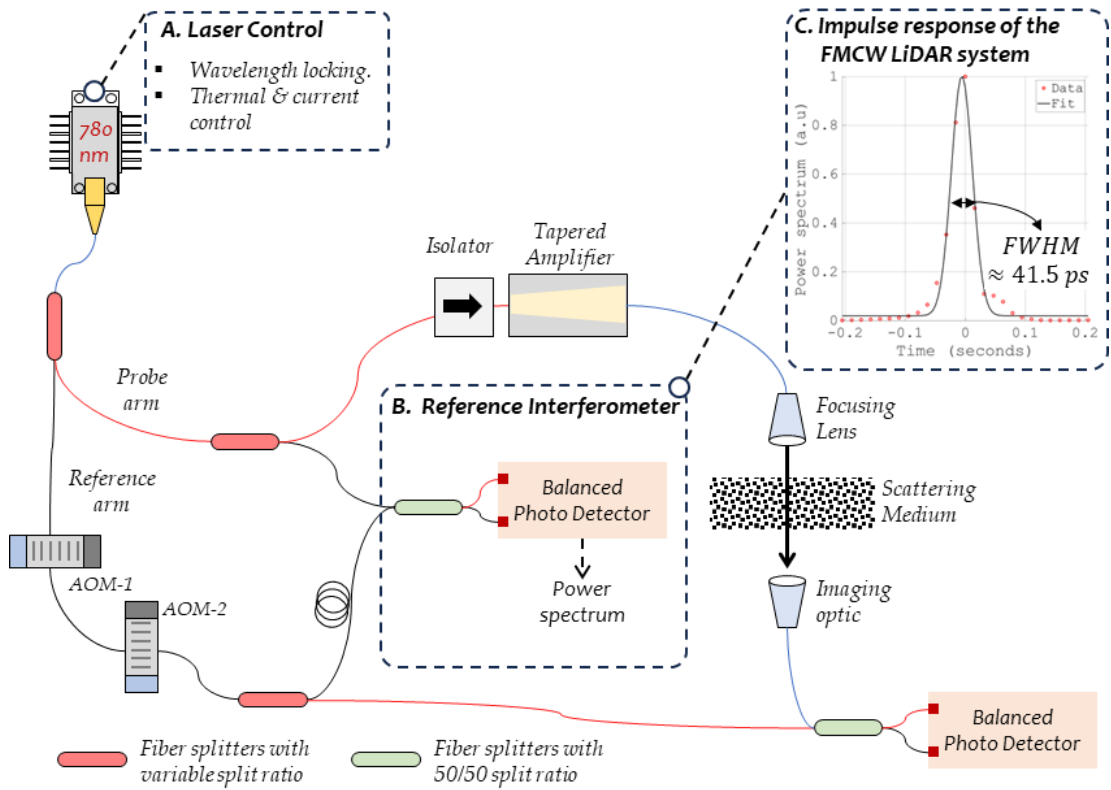

Figure S8: Full schematic of the FMCW LiDAR system.

To compensate for this non-linearity, we employ a cooperative modulation scheme[17,18] to correct these nonlinearities. The cooperative control scheme leverages the near instantaneous response of the optical frequency to changes in laser diode current. Furthermore, for small variations in current, the optical frequency sweep also varies linearly with current. Therefore, the previously measured non-linearities in the optical frequency sweep can be compensated by simultaneously applying a compensation waveform to the current modulation port. The correction waveform that is to be applied to the current modulation port is estimated by scaling $-v_n(t)$ using the current-to-wavelength conversion factor of the laser diode. For the laser we used in the experiments, this conversion factor was $0.68\ GHz/mA$. This estimated calibration waveform is then applied to the modulation input of the laser diode controller to correct the nonlinearity. The accuracy of the compensation is further improved by iteratively updating the compensation waveform to minimize the width of the beat frequency observed in the interferogram[16].

<u>Frequency Sweep Control and Synchronization:</u> The frequency sweep is initiated by a software trigger supplied to a four-quadrant current source (Keysight B2962A) programmed with the pre-calibrated waveform. The current source triggers an arbitrary waveform generator (Rigol Instruments, DG2102) that supplies the non-linearity correction waveform to the current modulation port of the laser controller (cooperative current control). The trigger is also used to synchronize the data captured on the high-speed digitizer (GaGe, RazorMax Express 16 CompuScope 16-Bit, 1 GS/s, PCIe Gen3 Digitizer) that digitizes the signal from the balanced photodetector.

After completing the frequency sweep using the pre-calibrated waveform, a two-stage a two-stage control scheme to lock the emission frequency of the laser engaged, as shown in Figure S9. In the first stage, a temperature PID control loop is engaged to lock the temperature of the laser to the vicinity of its original starting temperature. This PID controller, implemented on the PC in a MATLAB code, monitors the temperature of the laser by polling the thermistor values using a temperature controller (Slice – QTC). This PID controller drives the error signal between the desired and actual temperatures to zero by supplying a correction signal to TEC using the four-quadrant current source. However, merely locking the temperature of the laser back to its original value does not ensure that the emission frequency is restored to its original value, particularly when the temperature is tuned over a range of $\approx 20°\ C$. Therefore, an active frequency stabilization system is engaged to lock the wavelength of the laser back to the initial conditions.

This active frequency stabilization system utilizes the laser frequency control electronics comprising of a master laser operating at a fixed optical frequency, a high-speed beat note detector (Vescent Electronics, D2-260), and a Phase Frequency Detector (PFD) implemented on an offset phase lock servo (Vescent Electronics, D2-135). Here, the D2-260 records the interference between the laser emissions from the slave laser (used for probing the scattering medium) and the master laser, interfered using a 50:50 beam splitter. However, the high-speed beat detector can only measure the beat frequencies within the range of 250 MHz to 9 GHz. Therefore, it is essential for the temperature PID loop employed in the previous stage to bring the beat frequency to within 9 GHz of the desired frequency. This is a tighter constraint that increases the overall settling time of the PID loop.

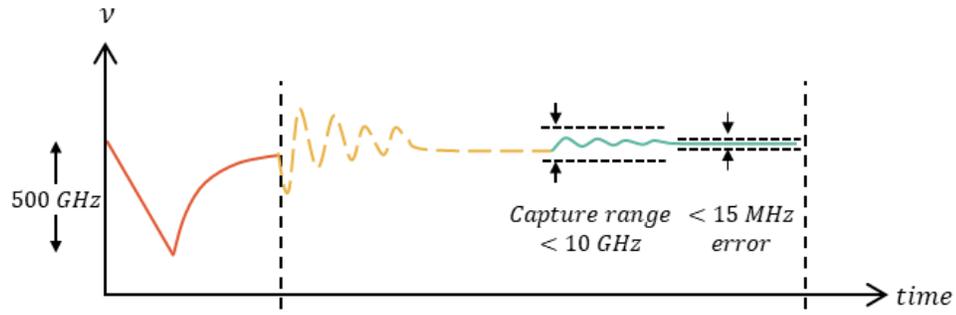

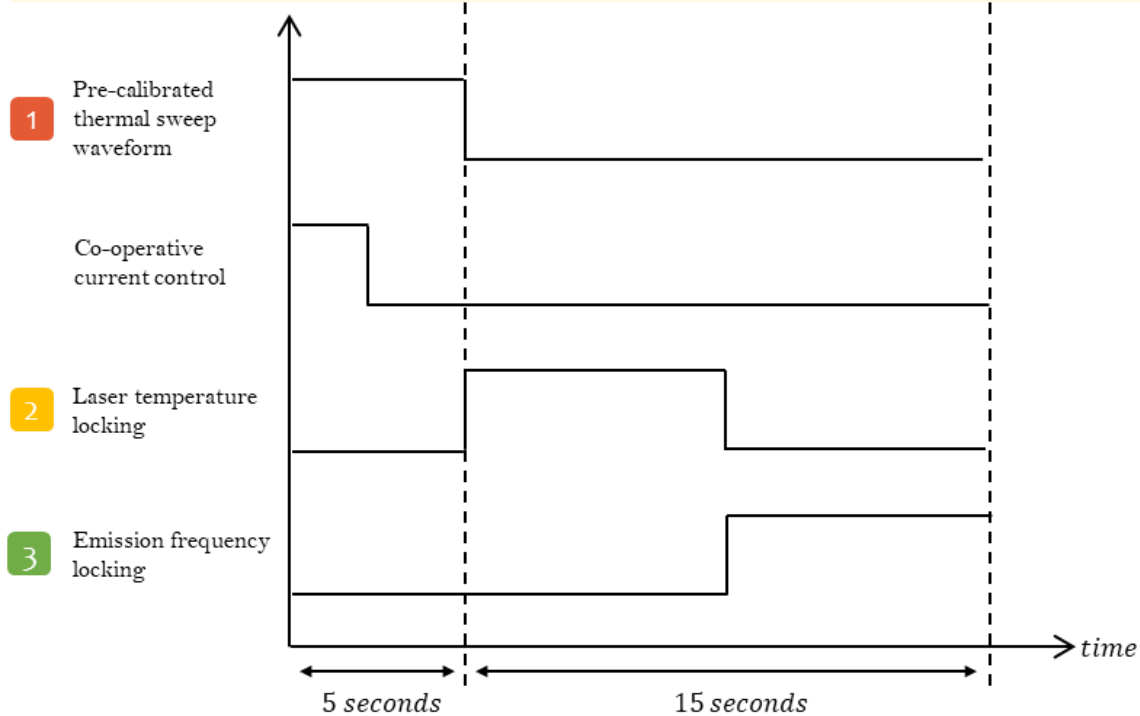

Figure S9: A) Schematic diagram of a typical optical frequency sweep waveform realized using the proposed approach B) Timing diagram of various control mechanisms

Once the frequency is within the capture range, the PFD compares the beat frequency from D2-260 to the desired frequency offset generated by the internal oscillator. The output voltage of the phase lock servo in the PFD mode is equal to $\pm 4\left(1 - \frac{f_{small}}{2f_{big}}\right)$, where $f_{small}$ and $f_{big}$ are the smaller and larger of the frequencies being compared. The error voltage generated by the PFD is accessed through an analog port on the D2-135 to build a PID controller on the PC. This PID controller drives the error voltage of the PFD to zero by supplying a correction signal to TEC using the four-quadrant current source. Each of these systems are sequentially engaged, as shown in Figure S9-B, to bring the emission frequency of the laser to within $15\ MHz$ of the master laser.

After convergence, the power spectrum of the measured interferogram is computed using with four non-overlapping subregions and a Blackman–Harris window function. A Gaussian fit to the measured impulse response indicates a standard deviation of 20 ps (see Figure S8-C). Additionally, the jitter the measured peaks were assessed by acquiring $\approx 430$ measurements spanning a duration of $\approx 4\ hrs$. The standard deviation and the mean absolute deviation (MAD) around the mean of these measurements were estimated to be $26\ ps$ and $19\ ps$, respectively.

Given these values, the system was operated at a temporal resolution of 45 ps, which is approximately twice the jitter. This also allowed averaging over more temporal windows to help mitigate speckle noise.

*Frequency stability of the master laser:* The repeatability of the starting frequency of the sweep is predicated in the stability of the emission frequency of the master laser. Therefore, the stability of the emission frequency of the master laser was examined by monitoring the beat frequency produced between the master laser and a reference laser locked to a Rb-absorption line. A Rubidium (Rb) Saturated Absorption Spectroscopy module (Vescent, D2-210) was used to stabilize the emission frequency of a $780\ nm$ tunable laser source by locking it to a hyperfine Rb-absorption line. The D2-210 is comprised of a Rb vapor cell, a temperature controller, and a balanced photo detector. The laser output, modulated at RF frequencies using a bias tee is transmitted through a vapor cell. In addition, the injection current of the laser is ramped up and down to modulate the instantaneous emission frequency of the laser. The sharp transitions in the saturated absorption spectrum, measured from the output of the balanced photodetector, are used to lock the laser frequency to the spectral line.

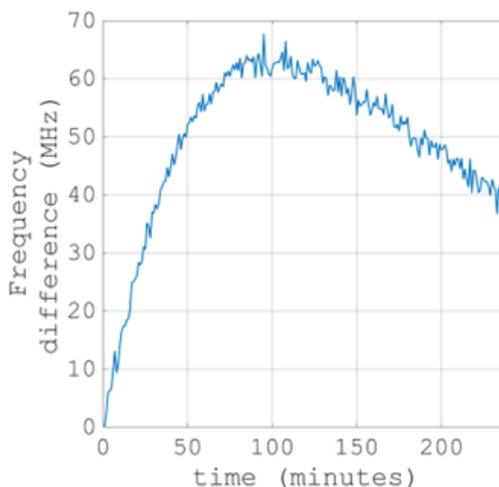

Figure S10: Drift in the emission frequency of the master laser as a function of time.

The beat frequency between the master laser and a reference laser was recorded on the high-speed beat note detector, whose signal was digitized using a high-speed digitizer. This signal was captured for $\approx 1s$ duration every minute for over a period of $6\ hrs$. The peak of the power spectrum of the recorded signal is plotted as a function of time (see Figure S10) to analyze the drift in the emission frequency of the laser. It can be noticed that the emission frequency of the laser takes $2\ hrs$ after startup to stabilize and then drifts at a rate of $8\ MHz/hr$. Since most of the acquisitions in this work finished within $2 - 4\ hrs$, this did not present any noticeable affects in the experiments and the acquired measurements.